\documentclass[11pt]{article}
\usepackage{latexsym}
\usepackage{epsf}

\def\be{\begin{equation}}
\def\ee{\end{equation}}
\def\bea{\begin{eqnarray}}
\def\eea{\end{eqnarray}}
\def\ba{\begin{array}}
\def\ea{\end{array}}
\def\nn{\nonumber \\}

\newcommand{\ul}{\underline}

\newcommand\tB{{\widehat B}}
\begin{document}
\begin{flushright}
IFUM-748-FT\\
\end{flushright}
\vspace{1truecm}

\centerline{\Huge On type II superstrings in bosonic backgrounds:}

\centerline{\Huge the role of fermions and T-duality}

\vspace{2truecm}

\begin{center}
 {\Large  Luca~Martucci\footnote{luca.martucci@mi.infn.it}
 and Pedro~J.~Silva\footnote{pedro.silva@mi.infn.it}}\\
\renewcommand{\thefootnote}{\arabic{footnote}}
\vspace{.5truecm}
Dipartimento di Fisica dell'Universit\`a di Milano,\\
Via Celoria 16, I-20133 Milano, Italy\\

\vspace*{0.5cm}

INFN, Sezione di Milano,\\
Via Celoria 16,
I-20133 Milano, Italy\\

\end{center}

\vspace{3truecm}

\centerline{\bf ABSTRACT}
\vspace{1truecm}
\noindent We derive the actions for type II Green-Schwarz strings up to second order in the fermions, for general bosonic backgrounds. We base our work on the so-called normal coordinate expansion. The resulting actions are $\kappa$-symmetric and, for the case of surviving background supersymmetries, supersymmetric. We first obtain the type IIa superstring action from the 11D supermem\-brane by double dimensional reduction. Then, by means of a generalization of T-duality we derive the type IIb superstring action. The resulting  actions are surprisingly simple and elegant.
\vspace{.5truecm}

\newpage

In this short letter we will present the type II superstring actions expanded in fermion coordinates up to second order in general supergravity bosonic backgrounds. In contrast to the usual situation where this expansion is written in a very unfriendly outcast, our result is elegant and economical. The action actually comes in terms of the ``natural" operators appearing in the supersymmetric transformation of the supergravity background fields. In the same footing, the well-known symmetries of these actions like $\kappa$-symmetry and supersymmetry are almost obvious from the expressions that we write here. This miracle, so to speak, is the result of using the so-called ``normal coordinate expansion" \cite{nc1} in the superfield formulation. Also, due to the close relation between supersymmetry and T-duality for world-volume theories, our action is also ``covariant'' under T-duality transformation. In fact, we will re-derive the T-duality rules obtaining a generalization of the Busher rules that again is elegant and simple. Nevertheless, here we point out that T-duality among string theory actions correspond to a much more complete test for string theory, since not only the zero modes are considered, but also all the excitations.

Due to the simplicity of the results, here we show the actions and leave the actual derivation for later\footnote{We use the Nambu-Goto form of the action, since it naturally arises from the double dimensional reduction of the M2-brane. Nevertheless is trivial to rewrite the action of the sting in the Polyakov form and indeed this is useful to perform T-duality}:
\bea
S_{(F)}=-\int{d^2\xi\sqrt{-\det(g)}}+{1\over
2}\int{d^2\xi\epsilon^{ij}b_{ij}} \;+\;i\int
d^2\xi\sqrt{-\det(g)}\;\bar{y}P_{(-)}\Gamma^i\tilde D_iy,
\label{f}
\eea
where for the type IIa case,
\bea
y=\left(y_+\;+\;y_-\right)\;\;\hbox{with}\;\;\Gamma^{11} y_{\pm}=\pm y_{\pm}\;\;\hbox{and}\;\; P_{(-)}=\hbox{${1\over 2}$}\left(1-\hbox{${1\over 2\sqrt{-g}}$}\epsilon^{ij}\Gamma_{ij}\Gamma^{11}\right)\;,
\label{fa}
\nonumber
\eea
\bea
\tilde{D}_m &=&D^{(0)}_m+W_m\;\;,\nn
&&D^{(0)}_m = \partial_m +\frac{1}{4}\omega_{mab}\Gamma^{ab}+
\frac{1}{4\cdot2!}H_{mab}\Gamma^{ab}\Gamma^{{11}}\;, \\
&&W_m = \frac18 e^\phi \left( \frac{1}{2!} {\bf F}^{(2)}_{ab}\Gamma^{ab}\Gamma_m\Gamma^{{11}}+ \frac{1}{4!}{\bf F}^{(4)}_{abcd}\Gamma^{abcd}\Gamma_m\right)\;.\nonumber
\label{fb}
\eea
and for the type IIb case,
\bea
y=\left(
\begin{array}{cc}
y_1\\
y_2\\
\end{array}\right)\;\;\; \hbox{with} \;\;\Gamma^{11} y_{(1,2)}=y_{(1,2)}\;\; \hbox{and}\;\; P_{(-)}=\hbox{${1\over 2}$}\left(1-\hbox{${1\over 2\sqrt{-g}}$}\sigma_3\otimes\epsilon^{ij}\Gamma_{ij}\Gamma^{11}\right)\;, \nonumber
\eea
\bea
\tilde{D}_m &=&\left(
         \begin{array}{cc}
      \hat D^{(0)}_{(1)m}&0 \\
          0&\hat D^{(0)}_{(2)m}
      \end{array}\right)\;\; +\;\;
     \left(
         \begin{array}{cc}
      0&\hat W_{(1)m} \\
          \hat W_{(2)m}&0
      \end{array}\right)  \;,\nn \nn
&&\hat D^{(0)}_{(1,2)m} = \partial_m +\frac{1}{4} \omega_{mab}\Gamma^{ab}\pm \frac{1}{4\cdot 2!}H_{mab}\Gamma^{ab}\;,\\
&&\hat W_{(1,2) m} = \pm\frac18 e^\phi \left({\bf F}^{(1)}_a\Gamma^a \mp \frac{1}{3!} {\bf F}^{(3)}_{abc}\Gamma^{abc}+
\frac{1}{2\cdot 5!}{\bf F}^{(5)}_{abcde}\Gamma^{abcde}\right)\Gamma_m \;, \nonumber
\eea
where in the expressions for $\hat D^{(0)}$ and $\hat W$, $\pm$ correspond to the label $(1,2)$. Also in all the above, $g$ is the 10D metric, $H=db$ and ${\bf F}^{(n)}$ are the RR field strength. Note that $\tilde D_m$ is precisely the operator appearing in the supersymmetry variation of the 10D gravitino, i.e. $\delta_\epsilon \psi_m=\tilde D_m \epsilon$ (see appendix for the explicit expressions, conventions and definitions).

In the following we first derive the type IIa action as a double dimensional reduction from the membrane of 11D. Second, we fully discuss the $\kappa$-symmetry transformation and its geometrical meaning from the point of view of the embedding of the 2d-theory into 10d space-time. Third, for the case of supergravity bosonic backgrounds that preserve some supersymmetry, we will give the
supersymmetric transformation of the world-sheet variables that correspond to symmetries of the action. Forth, we will derive the type IIb superstring action and give the explicit T-duality rules that interrelate the type IIa and type IIb theories. At last, we will give the $\kappa$-symmetry and supersymmetry transformations for the type IIb case. In the appendix we include conventions, definitions and the T-duality rules of Hassan.

To begin the actual derivation of the above results, our starting point is the supersymmetric membrane action in 11D, i.e.
\be
S=-\int{d^3\xi\sqrt{\det(-\textbf{G})}}- {1\over
6}\int{d^3\xi\varepsilon^{ijk}\textbf{B}_{kji}},
\ee
where $i=(0,1,2)$ and ($\textbf{G}$,$\textbf{B}$) are the pull-back to the membrane of the metric and 3-form super-fields of N=1 11D supergravity. Then, we borrow from \cite{gk1} the expression corresponding to the expansion in normal coordinates, in a general bosonic background, up to second order in the fermions,
\bea
&&S=S^{(0)}+S^{(2)},\label{mt}\\
&&S^{(0)}=-\int{d^3\xi\sqrt{-\det(G)}}-
{1\over 6}\int{d^3\xi\varepsilon^{ijk}b_{kji}},\label{m0}\\
&&S^{(2)}={i\over 2}\int d^3\xi\left\{\sqrt{-\det(G)}\left[ \bar{y}\Gamma^i\nabla_iy + \bar{y}T_i\Gamma^iy\right]+\right.\nonumber\\
&&\;\;\;\;\;\;\;\;\;\;\;\;\;\;\;\;\;\;\;\;\;\;\;\;\;\;\;\;\;\;\;\;\;\;\;\;\
\left.-\;\hbox{${1\over
2}$}\varepsilon^{ijk}\left[\bar{y}\Gamma_{ij}\nabla_ky
+\bar{y}T_i\Gamma_{jk}y\right]\right\},
\eea
where ($G,b$) are the bosonic pull-back of ($\textbf{G}$,$\textbf{B}$), $y$ is a real Majorana 32 component spinor,  $\Gamma$ are real gamma matrices, $\nabla_i$ is the usual pull-back of the spinor covariant derivative and $T_i$ is the pullback of $T_m={1\over 36}(\delta_m^n\Gamma^{pqr}+{1\over 8}\Gamma_m{}^{npqr})H_{npqr}$ (see the appendix for more details).
To obtain the type IIa string action we perform a double dimensional reduction of the above expression. Using the typical KK-ansatz for the 11D metric
\be
ds_{(11)}^2=e^{-\hbox{$2\over3$}\phi}ds_{(10)}^2+e^{\hbox{$4\over3$}\phi}(dz-C_mdx^m)^2\;,
\ee
after some straight forward but tedious algebra, we get the type IIa string action (\ref{f},\ref{fa}) in general bosonic backgrounds expanded up to second order in the fermions.

At this point, it is important to remark that this action is by construction $\kappa$-symmetric and supersymmetric (for the case of bosonic backgrounds with surviving supersymmetries). The above statement requires some attention and so, before deriving the type IIb string action, we pause and discuss these symmetries.

The $\kappa$-symmetry of the above actions can be understood as inherit from the superfield $\kappa$-symmetry of the membrane in 11D \cite{tow1}. Recall that only by fixing $\kappa$-symmetry we get the world-sheet supersymmetric theory transforming the 10D spinors into two dimensional ones. In our case, the resulting expanded action in normal coordinates, after double dimensional reduction, is written in terms of the pull-back of a covariant expansion of the 10d supergravity fields.
This expansion comes in terms covariant derivatives, curvature tensor, and the torsion. In principle the above type of Lagrangian contains too much information, related to the generalizations of concepts like extrinsic curvature, that are by no means related to a pure supersymmetric two-dimensional world-sheet theory. Therefore, the role of $\kappa$-symmetry is to eliminate all this redundant information obtaining a world-sheet theory. In other words $\kappa$-symmetry is the responsible to eliminate from the action all the geometrical information of the embedding that is not strictly speaking two-dimensional. In practice the expressions for the infinitesimal transformations, that leave our action invariant up to second order in $y$, are given by
\bea
\delta_\kappa y= (1+\Gamma_F)\kappa \;\;\;,\;\;\;
\delta_\kappa x^m= {i\over 2}\bar y \Gamma^m (1+\Gamma_F)\kappa\;\;\;,\;\;\;
\delta_\kappa A= \delta_\kappa x^m\partial_m A\;,
\label{akappa}
\eea
where $\Gamma_F=\hbox{${1\over 2\sqrt{-g}}$}\epsilon^{ij}\Gamma_{ij}\Gamma^{11}$ and $A$ is a general field of the supergravity background. These transformation rules are derivable using the normal coordinate formalism (see for example \cite{nc1}).

The supersymmetry transformations (again up to second order in $y$) are also derived within the same approach and are given by
\bea
\delta_\epsilon y =\epsilon \;\;\;,\;\;\;
\delta_\epsilon x^m= -{i\over 2}\bar y \Gamma^m \epsilon \;\;\;,\;\;\;
\delta_\epsilon A= \delta_\epsilon x^m\partial_m A\;,
\label{asusy}
\eea
where again $A$ is a general field of the supergravity background and the fermion $\epsilon$ is actually a killing spinor of the bosonic background, i.e. $\delta_\epsilon\psi_m=\delta_\epsilon\lambda=0$, where $\psi_m$ and $\lambda$ are the gravitino and the dilatino respectively (see appendix for explicit expressions). Therefore a translation along its orbit corresponds to a supersymmetry transformation of the Lagrangian.

Now that we understand the two more relevant symmetries of the above action, let us concentrate on the type IIb case. In order to obtain the corresponding action for the type IIb string, we perform a T-duality transformation, assuming the existence of a killing direction $x^9$. First of all, we rewrite the action (\ref{f},\ref{fa}) in the following form,
\bea 
S_{(IIa)}=-\int{d^2\xi\sqrt{-\det(G)}}+{1\over
2}\int{d^2\xi\epsilon^{ij}B_{ij}}
\eea
where
\bea
G_{mn}=g_{mn}-i\bar{y}\Gamma_{(m}\tilde D_{n)}y\;,\nn
B_{mn}=b_{mn}-i\bar{y}\Gamma^{11}\Gamma_{[m}\tilde D_{n]}y\;,
\eea
and we have used the fact that this action comes from the pull-back of the normal coordinate expansion of the vielbein and other supergravity fields obtaining, among other things, the pull-back of the covariant derivative of the fermionic variables $y$. Then, as is usual in this the context, T-duality is implemented with the help of a Lagrange multiplier $Z$ and a world-volume vector $V_i$ (taking the place of $\partial_i x^9$), by introducing a term of the form $\epsilon^{ij}\partial_iV Z_j$ into the action. Since the action formally has the same form as the bosonic string action, we only get the usual Busher rules\footnote{See \cite{enrique} and references therein.} once we solve for $V_i$.
\bea
\hat{G}_{ 99} &=& {1\over G_{99}}
\qquad\qquad\qquad\qquad
\qquad\qquad\qquad\qquad
\nonumber\\
\hat{G}_{ \tilde m \tilde n} &=& G_{\tilde m \tilde n}
- { G_{\tilde m 9} G_{\tilde n 9}
- B_{\tilde m 9} B_{\tilde n 9}
\over G_{ 99}}
\qquad\qquad\qquad\quad
\hat{G}_{\tilde m 9} ={ B_{\tilde m 9}
\over G_{99}}\\
\tB_{ \tilde m \tilde n}&=&B_{ \tilde m \tilde n}
-{B_{\tilde m 9} G_{\tilde n 9}-G_{\tilde m 9}
B_{\tilde n 9}\over G_{99}}
\qquad\qquad\qquad\quad
\tB_{\tilde m 9} ={ G_{\tilde m 9}
\over G_{ 99}}
\nonumber
\eea
where $\tilde m= 0,\ldots,8$. The important ingredient comes when we rewrite the fermionic part of the fields, ($G,B$) in terms of type IIb variables. At this point, because of the nature of normal coordinate expansion, we can use the transformation rules of Hassan (see the appendix) to obtain the desired result. The corresponding T-duality rules can be found in the appendix, here we write the final form of the transformed fields ($\hat G,\hat B$)\footnote{We should point out that the transformation rule of the dilaton is obtained with no need of including quantum corrections. Basically it is the curved space-time structure what provides a classical derivation for the dilaton T-duality.}:
\bea
\hat G_{mn}=\hat g_{mn}-i\bar{y}\Gamma_{(m}\tilde D_{n)}y\;,
\nn \hat B_{mn}=\hat b_{mn}-i\bar{y}\hat\Gamma\Gamma_{[m}\tilde
D_{n]}y \;,
\eea
where $\hat \Gamma= \sigma_3\otimes\Gamma^{11}$, ($y,\tilde D$) are define as in (\ref{fb}) and ($\hat g_{mn},\hat b_{mn}$) are transformed metric and NS field in the type IIb theory. Therefore, after some trivial manipulation we find that the expanded up to second order in the fermions type IIb fundamental string action in general bosonic backgrounds is given by equations (\ref{f},\ref{fb}).

Finally, it only remains to obtain the explicit form for the supersymmetry and $\kappa$-symmetry transformations for the type IIb action. To derive the rules for these symmetries we adapt an argument presented in \cite{tow2}. Let us begin with the derivation of the $\kappa$-symmetry rules for the type IIb superstring. We use the fact that originally the vector field $V_i$ obeys the constraint $\epsilon^{ij}\partial_i V_j=0$. At this point we know that the action is invariant under the transformation (\ref{akappa}). Therefore, since this constraint is invariant under $\kappa$-symmetry, the variation of $S$, in which we take $V_i$ as independent variable, should be proportional to $\epsilon^{ij}\partial_i V_j$.
Such a term can be easily isolated to be
\begin{equation}
\delta_\kappa S \sim \int \epsilon^{ij}\partial_{i}V_j[\delta_{\kappa} x^m b_{9m}-\frac i2 \bar y \Gamma^{ {11}}
(1+\Gamma_{F})\kappa] \ .
\end{equation}
Hence, to have a $\kappa$-invariant action, we must take
\begin{eqnarray}
\delta_\kappa Z=\frac i2 b_{9m}\bar y \Gamma^m(1+\Gamma_{F})\kappa-\frac i2 \bar y \Gamma^{{11}}
(1+\Gamma_{F})\kappa\ ,
\end{eqnarray}
that, written in IIb T-dual variables\footnote{$\Gamma_{F}$ dualize to $\hat\Gamma_{F}$, as can be read directly from the T-duality transformation rule of the action.}, takes the form
\begin{eqnarray}
\delta_\kappa Z=\frac i2 \bar y
(1+\hat\Gamma_{F})\kappa\ ,
\end{eqnarray}
where $\hat \Gamma_F=\hbox{${1\over 2\sqrt{-g}}$}\sigma_3\otimes\epsilon^{ij}\Gamma_{ij}\Gamma^{11}$ and the fermions $y$ are in the same convention used in (\ref{fb}). The other variables T-dualize trivially and it is now easy to complete the derivation of the transformation rules. Hence we get (up to second order in $y$),
\bea
\delta_\kappa y= (1+\hat \Gamma_F)\kappa \;\;\;,\;\;\;
\delta_\kappa x^m= {i\over 2}\bar y \Gamma^m (1+\hat \Gamma_F)\kappa\;\;\;,\;\;\;
\delta_\kappa A= \delta_\kappa x^m\partial_m A \;,
\label{bkappa}
\eea
where $A$ is a general field of the supergravity background.

The supersymmetry  transformations (again up to second order in $y$) are derived in the same footing, and are given by
\bea
\delta_\epsilon y =\epsilon \;\;\;,\;\;\;
\delta_\epsilon x^m= -{i\over 2}\bar y \Gamma^m \epsilon \;\;\;,\;\;\;
\delta_\epsilon A= \delta_\epsilon x^m\partial_m A\;,
\label{bsusy}
\eea
where again $A$ is a general field of the supergravity background
and we are assuming that the fermion $\epsilon$ is actually a killing spinor of the bosonic background.

In this letter we have provided a new form for the type IIa/b superstring actions, expanded up to second order in the fermions in general bosonic backgrounds.

There are two previous works on this subject. First \cite{yu}, where the type IIb Green-Schwarz superstring action is expanded in the N=2 10D superfield formulation. In this work, no discussion on supersymmetry, k-symmetry and T-duality is given. Second we have \cite{stelle}, which also starts from 11D as we do. Nevertheless, in that work the component expansion of the superfields is obtained using the so-called gauge completion expansion \cite{gc}. This method is based on a laborious comparison order-by-order between component supersymmetric transformations and superspace coordinate transformations. On the contrary, the normal coordinate expansion \cite{nc1} used here, gives a much more friendly framework with more powerful results (the action of the superstring is a good example).

We were able to cast the action in a ``covariant'' form for supersymmetry, $\kappa$-symmetry and T-duality. Also, we would like to stress that we consider only up to second order terms in the fermion expansion, not because of a limitation of the method but it was merely a choice that we took for brevity. The normal coordinate expansion can be carried out up to 32nd order if you wish, since it is defined by an iterative method with very simple rules. 

We have also given a generalization of the T-duality rules for these actions, that are in agreement with the rules of Hassan \cite{has,has1} (derived using pure supergravity arguments) and could be confronted with those found in \cite{stelle}. T-duality rules for the Green-Schwarz superstring in superspace formalism were studied in \cite{kulik} (see also \cite{ban}). Nevertheless, these studies are given in terms of superfields and therefore the structure of the T-duality rules for the components is implicit. It would be very interesting to compare both approaches by a normal coordinate expansion of their results.

As a last comment, it seems that the superstring actions founded here are well suited for studies on quantum properties like the beta function, the extension of the action to include couplings to the dilaton, etc...

\vspace{1cm}
\noindent
{\bf Acknowledgments}\\

We thank J. Gomis, M. Grisaru, D. Marolf and D. Zanon for useful discussions. This work was supported by INFN, MURST and by the European Commission RTN program HPRN-CT-2000-00131, in association with the University of Torino.

\vspace{1cm} \noindent{\bf \Large Appendix}
\vspace{.5cm}

{\em Conventions and definitions}
\vspace{.3cm}

For both 11D and 10D cases, we use for curved space indeces the letters $m,n,\ldots$ and $a,b,\ldots$ for flat space indices. We also underline explicit flat space indices (e.g., {\ul 0}, {\ul1}, etc.), to differentiate them from explicit curved indices. We take the metric to have signature $(-,+,...,+)$ and use the Clifford algebra\be
\{\Gamma^a,\Gamma^b\}=2\eta^{ab}\;,
\ee
where $\Gamma^a$ are real gamma matrices and $\eta^{ab}$ is the 10D Minkowski metric. We also set $\epsilon^{12\ldots}=1$, define $\Gamma^{11}= \Gamma^{\ul 0}\cdots\Gamma^{\ul 9}$ and use the notation $\Gamma_{a_1...a_n}=\Gamma_{[a_1}...\Gamma_{a_n]}$ denoting antisymmetrization with the corresponding weight $1/n!\;$; e.g. $\Gamma_{01} = \hbox{${1\over2}$}(\Gamma_0 \Gamma_1- \Gamma_1 \Gamma_0)$.

We use real Majorana anticommuting spinors of 32 components, denoted $y^\alpha$ or $\theta^\mu$. 
The conjugation operation is defined by,
\bea
\bar{y}=y^TC\;, \nn
\bar y_\beta = y^\alpha C_{\alpha\beta}\;,
\eea
where $T$ corresponds to transpose matrix multiplication; e.g. $y^\alpha C_{\alpha\beta}$ instead of $C_{\alpha\beta}y^\beta$, and $C_{\alpha\beta}$ is the antisymmetric charge conjugation matrix with inverse $C^{\alpha\beta}$.
The indices of a spinor and a bispinor $M^\alpha_{\;\;\beta}$ are lowered and 
raised via matrix multiplication by $C$ so that we have
\bea
C_{\alpha\beta}C^{\beta\gamma}=\delta^{\;\;\gamma}_{\alpha}\;, \nn
M_\alpha^{\;\;\beta}=C_{\alpha\gamma}M^\gamma_{\;\;\delta}C^{\delta\beta}\;, \nn
\bar{\theta}M\xi=\bar\theta_\alpha M^\alpha_{\;\;\beta} \xi^\beta =\theta^\alpha M_{\alpha\beta} \xi^\beta\;.
\eea
We take $C=\Gamma^{\ul{0}}$ and we use the usual Pauli matrices,
\[
\sigma^1 = \left( \ba{cc} 0 & 1 \\ 1 & 0 \ea \right)
\qquad \qquad \sigma^2 = \left( \ba{cc} 0 & -i \\ i & 0 \ea \right)
\qquad \qquad \sigma^3 = \left( \ba{cc} 1 & 0 \\ 0 & -1 \ea \right) \]
Also for the RR field strength we have
\[
{\bf F}^{(n)}_{m_1\cdots m_n}=n\partial_{[m_1} C^{(n-1)}_{m_2\cdots m_n]}-{n!\over 3!(n-3)!}H_{[m_1m_2m_3}C^{(n-3)}_{m_4\cdots m_n]}.
\]

\vspace{.5cm}
{\em Supergravity conventions}
\vspace{.3cm}

For the supergravity conventions and definitions we follow \cite{has} and \cite{antoine} .

The type IIa bosonic part of the action is given by
\bea
 S_{IIa} &=&
\frac{1}{2\kappa_{10}^2}\int d^{10} x \sqrt{-g}
    \Big\{
    e^{-2\phi} \big[
    R +4\big( \partial{\phi} \big)^{2}
    -\frac{1}{2 \cdot3!} H^2\big] + \nn
   && - \frac{1}{2\cdot 2!} ({\bf F}^{(2)})^2 - \frac{1}{2\cdot 4!} ({\bf F}^{(4)})^2 \Big\}
+ \frac{1}{4\kappa_{10}^2}\int b\wedge dC^{(3)}\wedge dC^{(3)}
\eea
and the supersymmetry transformations for the gravitino $\psi_m$ and dilatino $\lambda$ are
\bea
\delta\psi_m &=& \left[\partial_m +\frac{1}{4} \omega_{mab}\Gamma^{ab}+\frac{1}{4\cdot 2!}H_{mab}\Gamma^{ab}\Gamma^{\ul{\varphi}} \; + \right. \nn
&&\left. + \frac18 e^\phi \big( \frac{1}{2!} {\bf F}^{(2)}_{ab}\Gamma^{ab}\Gamma_m\Gamma^{\ul{\varphi}}+ \frac{1}{4!}{\bf F}^{(4)}_{abcd}\Gamma^{abcd}\Gamma_m\big)\right]\epsilon \ , \nn
\delta\lambda &=& \left[ \frac12 \left( \Gamma^m \partial_m\phi + \frac{1}{2\cdot 3!}H_{abc}\Gamma^{abc}\Gamma^{\ul{\varphi}}\right) \; +\right.\nn
&&+ \left. \frac{1}{8} e^\phi \left( \frac{3}{2!} {\bf F}^{(2)}_{ab}\Gamma^{ab}\Gamma^{\ul{\varphi}}+ \frac{1}{4!} {\bf F}^{(4)}_{abcd}\Gamma^{abcd}\right)\right] \epsilon \ .
\end{eqnarray}

The type IIb bosonic part of the action is given by
\bea
S_{IIb}&=&
\frac{1}{2\kappa_{10}^2}\int d^{10} x \sqrt{-g}
    \Big\{
    e^{-2\phi} \big[
    R +4\big( \partial{\phi} \big)^{2}
    -\frac{1}{2 \cdot3!} H^2\big] + \nn
   && - \frac{1}{2} ({\bf F}^{(1)})^2 - \frac{1}{2\cdot 3!} ({\bf F}^{(3)})^2 
- \frac{1}{4\cdot 5!} ({\bf F}^{(5)})^2  \Big\}+\cr
&&+ \frac{1}{4\kappa_{10}^2}\int db\wedge dC^{(2)}\wedge\big( C^{(4)}-\frac12 b\wedge C^{(2)}\big)\ ,
\eea
and the supersymmetry transformations for the gravitino $\psi_m$ and dilatino $\lambda$ are,
\bea
\delta\psi_{(1,2)m} &=& \left(\partial_m +\frac{1}{4} \omega_{mab}\Gamma^{ab}\pm\frac{1}{4\cdot 2!}H_{mab}\Gamma^{ab}\right)\epsilon_{(1,2)} +  \cr
&&+ \frac18 e^\phi \left(\pm {\bf F}^{(1)}_a\Gamma^a - \frac{1}{3!} {\bf F}^{(3)}_{abc}\Gamma^{abc}\pm
\frac{1}{2\cdot 5!}{\bf F}^{(5)}_{abcde}\Gamma^{abcde}\right)\Gamma_m \epsilon_{(2,1)}\;, \nn
\delta\lambda_{(1,2)} &=& \frac12 \left( \Gamma^m \partial_m\phi \pm\frac{1}{2\cdot 3!}H_{abc}\Gamma^{abc}\right)\epsilon_{(1,2)}\;+ \nn 
&&\hspace{2,6cm}+ \frac{1}{2} e^\phi \left( \mp  {\bf F}^{(1)}_{a}\Gamma^{a}+
\frac{1}{2\cdot 3!} {\bf F}^{(3)}_{abc}\Gamma^{abc}\right) \epsilon_{(2,1)}\; .
\eea
In the above expressions $2\kappa_{10}^2=(2\pi)^7 l_s^8 g_s^2$ and for the type IIb case we use the convention that the self duality constraint on ${\bf F}^{(5)}$ is imposed by hand at the level of the equations of motion and we have used the following convention for differential forms,
\[w^{(p)}=\hbox{${1\over p!}$}dx^{m_1}\wedge\cdots \wedge dx^{m_p} w_{m_p\cdots m_1}\ .\]

\vspace{.5cm}
{\em T-duality rules}
\vspace{.3cm}

Here we follow the Hassan formalism of \cite{has} for T-duality:
since we wish to apply T-duality along the 9th direction, let us introduce the following useful objects ( $\tilde m,\tilde n=0,\ldots,8$)
\begin{eqnarray}
&& \Omega=\frac{1}{\sqrt{g_{99}}}\Gamma^{11}\Gamma_9 \Rightarrow \Omega^2=-1\cr
&& E_{mn}=g_{mn}+b_{mn}\cr
&& (Q_{\pm})^m{}_n=\left(
\begin{array}{cc}
\mp g_{99} & \mp (g\mp b)_{9\tilde n}\\
 0         &   {\bf 1}_9 \\
\end{array} \right)\cr
&& (Q^{-1}_{\pm})^m{}_n=\left(
\begin{array}{cc}
\mp g_{99}^{-1} &  -g_{99}^{-1}(g\mp b)_{9\tilde n}\\
 0         &   {\bf 1}_9 \\
\end{array} \right)\ .
\end{eqnarray}
The T-duality rules for $E_{mn}$ are\footnote{Here and in the rest of this work, in ambiguous
cases, we will put a tilde over the transformed fields},
\begin{eqnarray}
\tilde E _{\tilde m\tilde n}&=& E _{\tilde m\tilde n}-E _{\tilde m 9}g_{99}^{-1}E _{9\tilde n}\cr
\tilde E _{\tilde m 9}&=& E _{\tilde m 9}g_{99}^{-1}\cr
\tilde E _{9 \tilde m }&=& -E _{9\tilde m }g_{99}^{-1}\cr
\tilde E _{9 9 } &=& g_{99}^{-1}\ .
\end{eqnarray}
For the transformation of the vielbein and the spinors, we will use the Hassan
conventions to avoid ambiguities\footnote{We recalled that there are two possible choices $e^m_{(\pm)a}$ for the transformed vielbein, related by a Lorentz transformation $\Lambda^b{}_{a}$.}.
The transformation rules for the vielbein are
\begin{eqnarray}
\tilde e^m_a\equiv e^m_{(-)a}= (Q_{-})^m{}_n e ^n_a \Rightarrow \tilde e ^a_m\equiv e^a_{(-)m}= (Q_{-}^{-1})^n{}_m e ^a_n\ .
\end{eqnarray}
We will also need the alternative transformed vielbein
\begin{eqnarray}
 e^m_{(+)a}= (Q_{+})^m{}_n e ^n_a=\Lambda^b{}_{a} e^m_{(-)b} \Rightarrow e^a_{(+)m}=
(Q_{+}^{-1})^n{}_m e ^a_n=e^b_{(-)m}\Lambda_b{}^a\ .
\end{eqnarray}
Therefore going from IIa to IIb, we have:
\begin{eqnarray}
y_+ &=& y_1 \Rightarrow  \bar y_+ = \bar y_1\cr
y_- &=& -\Omega y_2 \Rightarrow  \bar y_- = \bar y_2 \Omega\cr
&&\cr
D^{(0)}_m y_+ &=& (Q_{+}^{-1})^n{}_m(\hat D^{(0)}_{(1)n} y_1)\cr
D^{(0)}_m y_- &=& -\Omega(Q_{-}^{-1})^n{}_m(\hat D^{(0)}_{(2)n} y_2)\cr
W_m y_+ &=& -\Omega(Q_{-}^{-1})^n{}_m(\hat W_{(1)n} y_1)\cr
W_m y_- &=& (Q_{+}^{-1})^n{}_m(\hat W_{(2)n} y_2)\nonumber
\end{eqnarray}
Conversely, going from IIb to IIa we have
\begin{eqnarray}
y_1 &=& y_+ \Rightarrow  \bar y_1 = \bar y_+\cr
y_2 &=& \Omega y_- \Rightarrow  \bar y_2 = -\bar y_- \Omega\cr
&&\cr
\hat D^{(0)}_{(1)m} y_1 &=& (Q_{+}^{-1})^n{}_m (D^{(0)}_n y_+)\cr
\hat D^{(0)}_ {(2)m}y_2 &=& \Omega(Q_{-}^{-1})^n{}_m(D^{(0)}_n y_-)\cr
\hat W_{(1)m} y_1 &=& \Omega(Q_{-}^{-1})^n{}_m(W_n y_+)\cr
\hat W_{(2)m} y_2 &=& (Q_{+}^{-1})^n{}_m(W_n y_-)\nonumber
\end{eqnarray}


\end{document}